# Current transport in Ni Schottky barrier on GaN epilayer grown on free standing substrates

Giuseppe Greco [1], Patrick Fiorenza [1], Emanuela Schilirò [1], Corrado Bongiorno [1], Salvatore Di Franco [1], Pierre-Marie Coulon [2], Eric Frayssinet [2], Florian Bartoli [2], Filippo Giannazzo [1], Daniel Alquier [3], Yvon Cordier [2], Fabrizio Roccaforte [1]

[1] Consiglio Nazionale delle Ricerche – Istituto per la Microelettronica e Microsistemi (CNR-IMM), Catania, Italy

[2] Université Côte d'Azur, CNRS, CRHEA, Valbonne, France

[3] Université Tours and GREMAN, Tours (France)

**Abstract**

In this paper, the Ni Schottky barrier on GaN epilayer grown on free standing substrates has been characterized. First, transmission electrical microscopy (TEM) images and nanoscale electrical analysis by conductive atomic force microscopy (C-AFM) of the bare material allowed visualizing structural defects in the crystal, as well as local inhomogeneities of the current conduction. The forward current-voltage (I-V) characteristics of Ni/GaN vertical Schottky diodes fabricated on the epilayer gave average values of the Schottky barrier height of 0.79 eV and ideality factor of 1.14. A statistical analysis over a set of diodes, combined with temperature dependence measurements, confirmed the formation of an inhomogeneous Schottky barrier in this material. From a plot of $\Phi_B$ versus n, an ideal homogeneous barrier close to 0.9 eV was estimated, similar to that extrapolated by capacitance-voltage (C-V) analysis. Local I-V curves, acquired by means of C-AFM, displayed the inhomogeneous distribution of the onset of current conduction, which in turn resembles the one observed in the macroscopic Schottky diodes. Finally, the reverse characteristic of the diodes fabricated in the defects-free region have been acquired at different temperature and its behaviour has been described by the thermionic field emission (TFE) model.

**1. Introduction**

Gallium Nitride (GaN) is considered an excellent semiconductor material to fulfill the requirements of the next generation of high-power and high-frequency devices with a high energy efficiency [1,2]. In fact, thanks to its large band gap (3.4 eV), high critical field (> 3 MV/cm) and high electron mobility in

AlGaN/GaN heterostructures (> 1000 cm$^2$V$^{-1}$s$^{-1}$), GaN exhibits theoretical figures of merit that are superior not only to Si but, in principle, also to the most mature silicon carbide (4H-SiC) [3]. However, there are still several open scientific and technological issues that characterize GaN technology, which are mainly related to fundamental material aspects, devices processes and reliability (e.g., metal contacts, dielectrics, etc.) [4].

Till now, the lack of large area bulk material of good quality has restricted GaN technology to lateral devices architectures on foreign substrates. In fact, while today a variety of lateral high electron mobility transistors (HEMTs), both on Si and SiC substrates, have already reached the market [5], vertical GaN power devices (Schottky, MOSFETs, etc.) are still limited to proofs of concept [6,7,8,9,10]. However, the vertical topology is highly desired for power electronics to further reduce the chip-size and simplify the packaging design. In this context, the metal/semiconductor interface represents the most important part of the Schottky diode, as it rules the entire electrical behavior of the device. Hence, studying the electrical properties of Schottky contacts in relation to the properties of homoepitaxial GaN epilayers is an important topic in this emerging field of research. However, the absence of lattice parameter mismatch makes the GaN epilayer properties more dependent with the quality of the GaN substrate, which motivates the present study. In previous papers, the high leakage current in Schottky diodes on bulk GaN was correlated with the presence of structural defects in the epilayer [11,12,13]. Moreover, the mechanisms of current transport dominating the electrical behaviour at the metal/GaN interface have been recently discussed [14,15]. New metallization schemes or advanced field plate structures have been proposed to reduce the leakage current in vertical Schottky diodes on homoepitaxial GaN epilayers. [16,17,18].

In the present paper, Ni Schottky diodes have been fabricated on a homoepitaxial layer grown onto a free standing GaN substrate. Then, the Schottky diodes have been electrically characterized in both forward and reverse bias at different temperatures. Differently from previous literature work, here a correlation between macroscopic electrical analyses on diodes with nanoscale electrical characterization on the epilayer is reported. The cross correlation of these measurements enabled to explain the Schottky barrier inhomogeneity and identify the dominant carrier transport mechanisms at the metal/GaN interface.

## 2. Experimental

The experiments were carried on 5 μm thick n-type GaN epitaxial layer grown onto heavily doped ($N_D$ = 1×10$^{19}$ cm$^{-3}$) free standing GaN substrates provided by Saint Gobain Lumilog [19]. The GaN epitaxial layer had a nominal net doping concentration N$_D$-N$_A$ = 2×10$^{16}$ cm$^{-3}$, as determined by Hg-probe capacitance-voltage (C-V) measurements. The GaN substrate was grown by hydride vapor phase epitaxy (HVPE), while metalorganic chemical vapor phase epitaxy (MOVPE) has been used to grow the epitaxial layer. X-ray diffraction (XRD) analyses gave a full width at half maximum of the rocking curves taken around GaN

(002) FWHM ~ 150 arsec, thus revealing a good average crystallographic quality of the layer. From XRD, the mean dislocation density was estimated to be inferior to $10^7/cm^2$.

A large area ohmic contact was created on the wafer back-side by a $Ti_{(15nm)}/Al_{(200nm)}/Ni_{(50\ nm)}/Au_{(50\ nm)}$ multilayer, followed by an annealing at 850 °C. The front-side of the wafer was subjected to a wet cleaning in diluted hydrofluoric acid prior to Schottky metal deposition. Then, circular diodes with radius of 50 μm, have been defined by standard photolithography, metal deposition and lift-off technique, using a $Ni_{(50nm)}/Au_{(50nm)}$ bilayer as Schottky electrode. The electrical characterization of the diodes has been carried out on a Karl–Suss MicroTec probe station equipped with a parameter analyser. Nanoscale resolution morphological and electrical characterization of the bare GaN surface was performed by conductive atomic force microscopy (C-AFM) using a DI3100 equipment with Nanoscope V controller electronics. Doped diamond coated Si tips with curvature radius of 10 nm and cantilever spring constant of 80 N/m were used for local electrical measurements. Force-distance curves have been acquired to properly set the deflection set-point, in order to minimize the applied force. Plan view transmission Electron Microscopy (TEM) analysis has been performed using a 200 kV JEOL 2010F microscope.

## 3. Results and discussion

### A. Nanoscale assessment of the epilayer quality

The GaN epitaxial layers have been firstly investigated by plan view TEM analysis. In this analysis, the GaN epitaxial layer appears mostly as free from defects (Fig.1a), with the presence of regions where few defects can be observed (Fig.1b). The pits visible in the plan-view TEM image in Fig. 1b can be associated to the presence of threading dislocations [20,21].

Under the morphological point of view, a very smooth surface of the epilayer was observed by AFM, with a root mean square roughness (RMS) of 0.1 nm. On the other hand, interesting insights were gained by the C-AFM measurements. These measurements have been performed by applying a negative bias of -2.5 V to the back of the sample and collecting the current to the conductive tip on the GaN surface. Fig1c and Fig.1d show the C-AFM maps acquired on a 10×10 μm$^2$ scan area, on the surface of the GaN epitaxial layer. While the defect-free region exhibits a very uniform current map, few spots with a higher conductivity are visible in the region containing defects. This observation demonstrates the electrical activity of the defects already observed by TEM (i.e. threading dislocations), which act as preferential vertical leakage paths in the GaN epitaxial layer [22].The defects density was typically in the order of $10^5$-$10^6$ cm$^{-2}$. However, some more defective parts of the sample were identified, in which the defects density was higher (i.e. $1\times10^7$ cm$^{-2}$). This non uniformity is due to the process used to grow the free-standing GaN substrate which results in the formation of defect clusters [19].

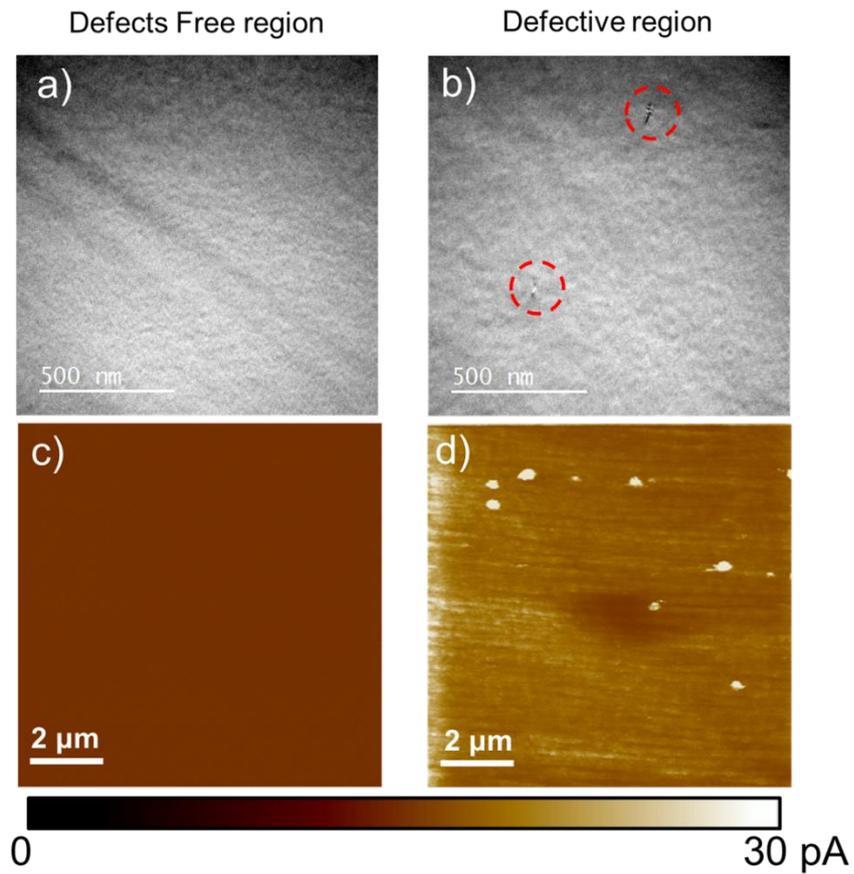

**Figure 1.** (a) TEM plan-view images of the GaN epitaxial layer in a "defect-free region" (a) and in a "defective region" (b). The red circles indicate the observed defects. C-AFM maps, acquired on a 10×10 µm² area of the GaN epilayer, in a "defect free region" (c) and "defective region" (d).

*B. Electrical characterization of Schottky diodes*

Statistical electrical characterizations on the diodes fabricated on this material have been carried out by means of current-voltage measurements in forward bias. Fig. 2a shows the current density vs voltage (J-V) characteristics measured at room temperature. The forward J–V curves display a linear behaviour in semi-log scale, with the current density that first increases up to 0.6 V and then saturates, due to the onset of the series resistance region

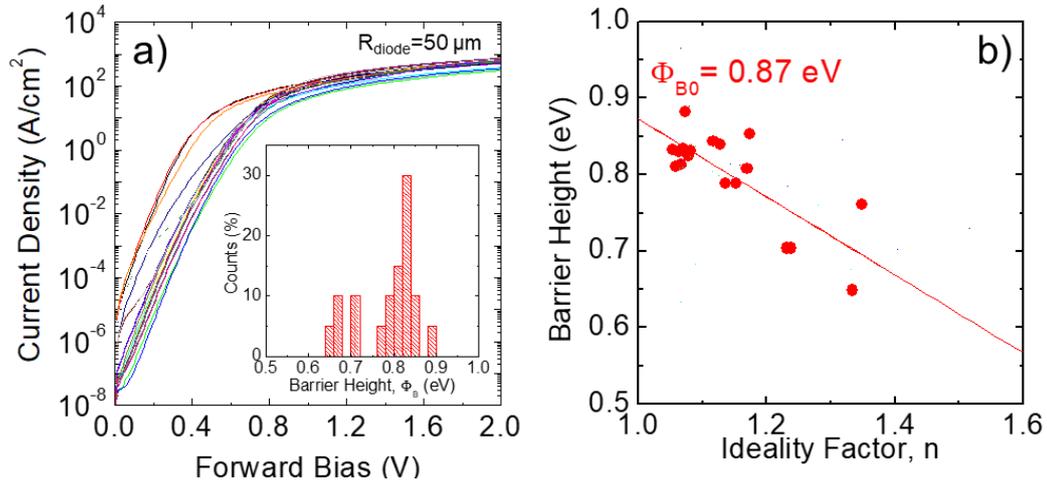

**Figure 2.** (a) Forward J-V curves acquired in different Schottky diodes and histogram of the extrapolated Schottky barrier height $\Phi_B$ values (inset). (b) Plot of the Schottky barrier height $\Phi_B$ as a function of the ideality factor n.

The electrical behaviour of the diodes was described by the Thermionic Emission (TE) mechanism, where the current density in the linear region is expressed by:

$$J = J_S \, exp\left(\frac{qV}{nkT}\right) \qquad \text{Eq. 1}$$

with the saturation current density $J_S$ given by

$$J_S = A^* T^2 \, exp\left(-\frac{q\phi_B}{kT}\right) \qquad \text{Eq. 2}$$

where A* is the Richardson constant (26.9 A/(cm$^2$K$^2$), T is the absolute measurement temperature, k is the Boltzmann constant, and $\Phi_B$ and n are the Schottky barrier height and the ideality factor, respectively. The inset of Fig. 2a reports a histogram of the barrier height values $\Phi_B$ extrapolated from the experimental J-V curves. As can be seen, the statistical distribution exhibits an average barrier value close to 0.79 eV, with some diodes showing a lower barrier close to 0.7 eV. Such distribution clearly indicates the presence of an inhomogeneous Schottky barrier [23]. Schottky contacts typically show a linear correlation between the n and $\Phi_B$, which depends on the degree of inhomogeneity of the barrier [24,25]. In particular, correlating n and $\Phi_B$ enables to extrapolate the values of the ideal homogenous barrier, corresponding to the barrier value obtained for n=1. Fig. 2b plots the values of the barrier height $\Phi_B$ as a function of the ideality factor n extrapolated from the linear fit of the semilog J-V curves. As can be seen, these two parameters exhibit a linear correlation, i.e. the higher barrier values, the better the ideality factor of diodes and vice versa. By a linear fit of the data, from the extrapolation of $\Phi_B$ at n=1, an "ideal" homogeneous barrier of $\Phi_{B0}$= 0.87 eV has been obtained, clearly higher than the average ones (0.79 eV).

Then, the forward J-V characteristics have been measured also at different temperatures in the range between 25°C and 125°C. As can be seen in Fig. 3a, the measured current on a Schottky diode increases with increasing the temperature. By applying again the TE model, it was possible to extract the values of the ideality factor n and of the Schottky barrier $\Phi_B$ at each temperature, which are reported in Fig.3b. In particular, the ideality factor of the diode decreases with increasing the temperature, while the barrier height increases with temperature. Similarly to what already shown for the room temperature analyses on different diodes, also from these data it was possible to generate a plot of the barrier height $\Phi_B$ versus ideality n determined at each temperature on a single diode. Interestingly, also in this case a linear correlation between these two parameters can be deduced, which confirm the inhomogeneity of the barrier [26]. The ideal homogeneous barrier extrapolated at n=1 was estimated to be 0.9 eV, very close to the value found by the $\Phi_B$ vs n plot determined at room temperature (Fig. 2b). The same barrier height value of 0.9 eV could be determined by C-V analysis (see Fig 3d) from the x-axis intercept of the $1/C^2$ vs V, shown as inset of Fig.3d. From the slope of the linear fit of the $1/C^2$ vs V plot, a donor concentration $N_D=2.5\times10^{16}$ cm$^{-3}$ could be determined, which is in good agreement with the nominal doping of the epilayer and the Mercure-probe C-V performed on the sample before process.

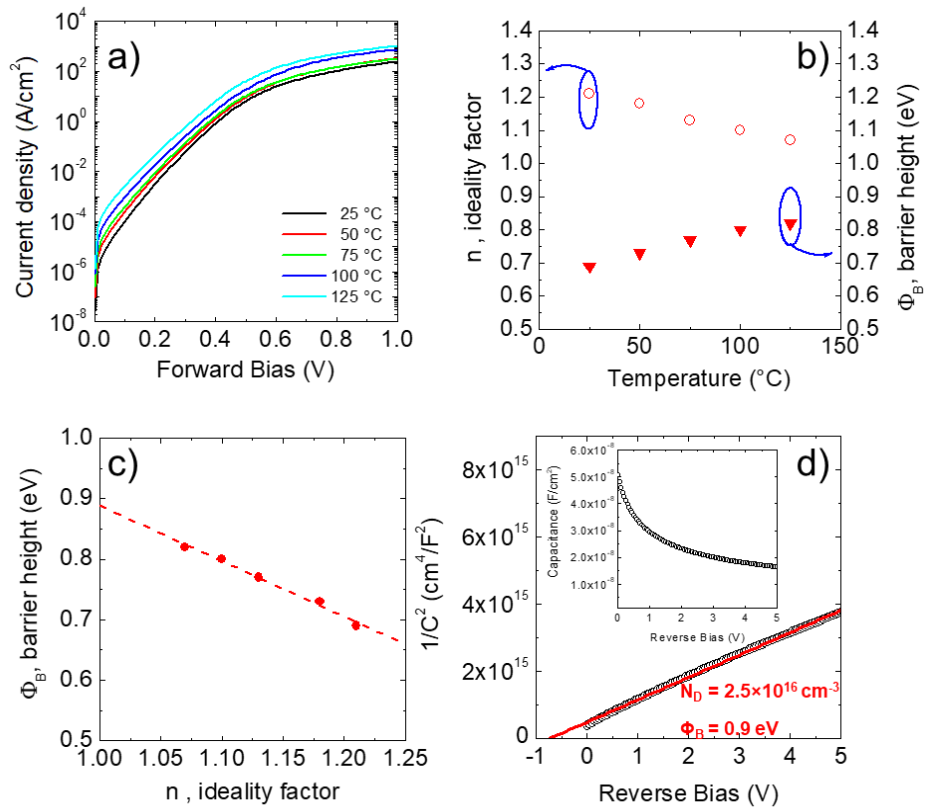

**Figure 3.** (a) Forward J-V curves acquired on a Ni/GaN Schottky diode at different temperatures; (b) Ideality factor n and Schottky barrier height $\Phi_B$ as a function of the temperature T; (c) Plot of the Schottky barrier height $\Phi_B$ as a function of the ideality factor n; (d) the $1/C^2$ versus reverse bias extrapolated by the C-V curves of the diode (inset).

In order to get additional information on the behaviour of the barrier at the nanoscale, local current voltage measurements have been carried out by C-AFM on the bare surface of the GaN epilayer. Fig. 4a reports an arrays of 25 local current-voltage (I-$V_{tip}$) acquired with a spacing of 10 µm. As can be seen, the experimental I-V curves exhibit a certain dispersion, which confirms the local inhomogeneity of the current conduction, i.e. of the barrier. From these curves, an onset voltage $V_{ON}$ was defined as the tip bias ($V_{tip}$) corresponding to a current of $5\times10^{-3}$ pA. Notably, the histogram of the $V_{ON}$ values, reported in Fig. 4b, shows the presence of two different $V_{ON}$ distributions, thus graphically resembling the same inhomogeneity observed at macroscopic level (see inset of Fig. 2a).

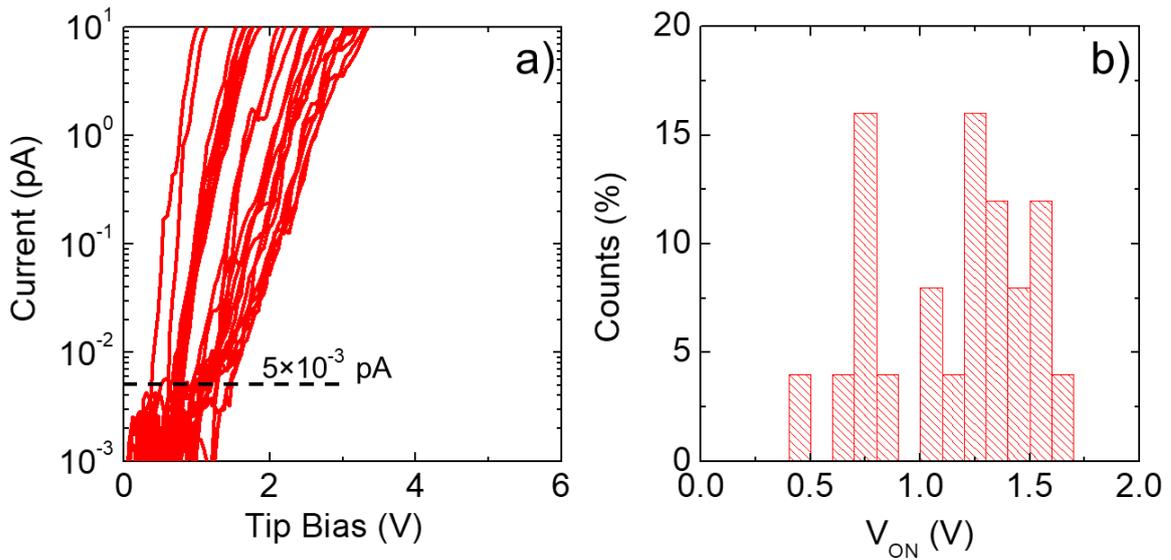

**Figure 4.** (a) Arrays of local current-voltage (I-$V_{tip}$) curves measured by C-AFM on the bare surface of the GaN epilayer; (b) Histogram of the onset voltage of the current conduction ($V_{ON}$) extrapolated from the local I-V curves at the current of $5\times10^{-3}$ pA.

The reverse J-V characteristics of the diodes fabricated in the defect-free region have been studied at different temperatures (see Fig. 5a). In particular, the reverse current of the diodes measured at 25 °C slightly increases with increasing of the reverse applied bias up to reach the value $4.2\times10^{-6}$ A/cm$^2$ at $V_R = -10$ V. At the same bias, an increase of the temperature at 75°C or 125°C results into an increase of the reverse leakage current to $2.2\times10^{-4}$ A/cm$^2$ or $1.3\times10^{-3}$ A/cm$^2$, respectively.

In wide band gap semiconductors, the thermionic Field Emission theory has been often used to explain the reverse J-V characteristic in Schottky Diodes [27,28,29]. The experimental J-V curves (at 25, 75 and 125°C) are reported together with those calculated using the thermionic field emission model (TFE) [30]. Accordingly, the current density is expressed by

$$J_{TFE} = A^*T^2 \sqrt{\frac{q\pi E_{00}}{kT}} \sqrt{V_R + \frac{\overline{\overline{\Phi_B}}}{\cosh\left(\frac{qE_{00}}{kT}\right)^2}} \exp\left(-\frac{\overline{\overline{\Phi_B}}}{E_0}\right) \exp\left(\frac{qV_R}{kT} - \frac{qV_R}{E_0}\right) \quad \text{(Eq. 1)}$$

with $E_0 = E_{00} \coth\left(\frac{E_{00}}{kT}\right)$, $E_{00} = \frac{qh}{4\pi}\sqrt{\frac{N_D}{m^*\epsilon}}$ and with, m* and ε the effective mass of electron and dielectric constant of GaN.

The image force lowering effect has been considered by using the following expression for the barrier height $\overline{\overline{\Phi_B}}$.

$$\overline{\overline{\Phi_B}} = \Phi_B - \left[\frac{q^3 N_D}{8\pi^2 \epsilon^3}(V_{bi} - V_R)\right]^{1/4} \quad \text{(Eq. 2)}$$

As can be seen, the reverse leakage current curves calculated using the TFE model are in good agreement with the experimental ones. However, at room temperature, at low applied bias a discrepancy between the theoretical TFE curve and the experimental data is observed. A possible explanation is the occurrence of different current transport mechanisms, which coexist with TFE. A similar conclusion has been reported in literature to model the electrical behaviour of Schottky diodes on GaN bulk in forward and reverse configuration [14,31].

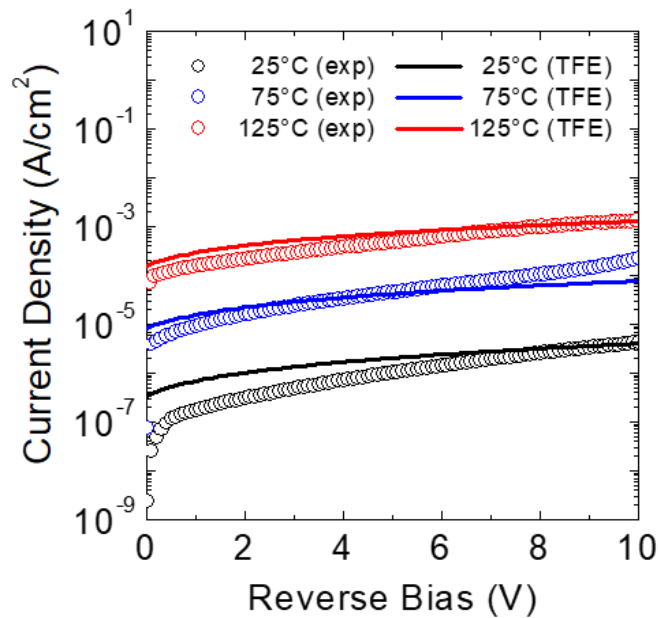

**Figure 5.** Reverse experimental J-V curves acquired on a Ni/GaN Schottky diode at 25, 75 and 125 °C (scatter) and the respectively theoretically J-V curves calculated according TFE model.

**Conclusions**

In summary, the Ni Schottky barrier has been characterized on GaN epilayer grown on bulk substrates. The forward characteristics of the diodes were described by thermionic emission model, with an average barrier height of 0.79 eV and ideality factor of 1.14. A statistical analysis over a set of diodes, combined with temperature dependence electrical measurements, demonstrated that an inhomogeneous Schottky barrier has formed. Nanoscale local currents-voltage measurements on the material displayed the inhomogeneous distribution of the onset of current conduction, in agreement with the inhomogeneity observed in the macroscopic diodes. The reverse leakage current of the diodes fabricated in the defects-free region could be described by the thermionic field emission (TFE) model, as already observed in other wide band gap semiconductors. The cross-correlation of macroscopic and nanoscale measurements presented in this paper is a powerful tool not only to get insights on the barrier inhomogeneity in relation to the quality of the epilayer, but also to better understand and optimize the behavior of vertical devices on free standing GaN.

**Acknowledgments**

This work has been carried out in the framework of the European Project GaN4AP (Gallium Nitride for Advanced Power Applications). The project has received funding from the Electronic Component Systems for European Leadership Joint Undertaking (ECSEL JU), under grant agreement No.101007310. This Joint Undertaking receives support from the European Union's Horizon 2020 research and innovation programme, and Italy, Germany, France, Poland, Czech Republic, Netherlands.

**References**

[1] F. Roccaforte and M. Leszczyński, Nitride Semiconductor Technology: Power Electronics and Optoelectronic Devices, 1st ed., Wiley-VCH Verlag GmbH & Co. KGaA, 2020.
[2] F. Roccaforte, P. Fiorenza, R. Lo Nigro, F. Giannazzo, & G. Greco. (2018). Rivista Del Nuovo Cimento, 41(12), 626–673
[3] F. Roccaforte, P. Fiorenza , G. Greco, R. Lo Nigro, F. Giannazzo, A. Patti, M. Saggio, Phys. Status Solidi a 211, 2063 (2014).
[4] F. Roccaforte, P. Fiorenza, G. Greco, R. Lo Nigro, F. Giannazzo, F. Iucolano, M. Saggio, Microelectr. Eng. 187–188 (2018) 66–77.

[5] K.J. Chen, O. Häberlen, A. Lidow, Chun lin Tsai, T. Ueda, Y. Uemoto, Y. Wu, IEEE Trans. Electron Devices 64, (2017) 779.
[6] Y. Sun, X Kang, Y Zheng, J Lu, X Tian, K Wei, H Wu, W. Wang, X. Liu and G. Zhang, Electronics, vol. 8, no. 5, pp. 575, (2019)
[7] J. Hu, Y. Zhang, M. Sun, D. Piedra, N. Chowdhury and T. Palacios, Mater. Sci. Semicond. Process.,78, 75-84,( 2018).
[8] Y. Zhang, M. Sun, Z. Liu, D. Piedra, J. Hu, X. Gao, and T. Palacios Appl. Phys. Lett. 110, 193506 (2017) .
[9] F. Roccaforte, F. Giannazzo, A. Alberti, M. Spera, M. Cannas, I. Cora, B. Pécz, F. Iucolano, and G. Greco, Mater. Sci. Semicond. Process. 94, 164 (2019)
[10] K. Fu, H. Fu, X. Huang, T.-H. Yang, C.-Y. Cheng, P. R. Peri, H. Chen, J. Montes, C. Yang , J. Zhou, X. Deng, X. Qi, D. J. Smith, S. M. Goodnick And Y.Zhao, IEEE Journal of the Electron Devices Society, vol. 8, pp. 74-83, 2020
[11] L. Sang, B. Ren, M. Sumiya, M. Liao, Y. Koide, A. Tanaka, Y. Cho, Y. Harada, T. Nabatame, T. Sekiguchi, S. Usami, Y. Honda, and H. Amano, Appl. Phys. Lett. 111, 122102 (2017).
[12] S. Li , B. Ercan , C. Ren , H. Ikeda, and S. Chowdhury, IEEE Trans. Electron. Dev. 69, 4206 (2022).
[13] G. Greco, F. Giannazzo, P. Fiorenza, S. Di Franco, A. Alberti, F. Iucolano, I. Cora, B. Pecz, F. Roccaforte, Phys. Status Solidi A, 215) (2018), 1700613.
[14] P.V. Raja, C. Raynaud, C. Sonneville, A.J.E. N'Dohi, H. Morel, L.V. Phung, T.H. Ngo, P.D. Mierry, E. Frayssinet, H. Maher, J. Tasselli, K. Isoird, F. Morancho, Y. Cordier, D. Planson, Microelectron. J. 128, (2022) 1055.
[15] A.Sandupatla, S. Arulkumaran, G.I. Ng, K. Ranjan, M. Deki, S. Nitta Y. Honda, H. Amano, Appl. Phys. Exp. 13, 074001 (2020).
[16] Z. Shi, X. Xiang, H. Zhang, Q. He, G. Jian, K. Zhou, X. Zhou, C. Xing, G. Xu, S. Long, Semicond. Sci. Technol. 37 (2022) 065010.
[17] X. Liu, F. Lin, J. Li, Y. Lin, J. Wu, H. Wang, X. Li, Shuangwu Huang, Q. Wang , H.C. Chiu, H.C. Kuo, IEEE Trans. Electron. Dev. 69, 1938 (2022).
[18] V. Maurya, J. Buckley, D. Alquier, H. Haas, M.R. Irekti,T. Kaltsounis, M. Charles, N. Rochat, C. Sonneville,V. Sousa, Microelectronic Engineering 274 (2023) 111975.
[19] T.H. Ngo, R. Comyn, E. Frayssinet, H. Chauveau, S. Chenot, B. Damilano, F. Tendille, B. Beaumont, J.P. Faurie, N. Nahas, Y. Cordier, Journal of Crystal Growth 552 (2020) 125911,
[20] G. Greco, F. Iucolano, C. Bongiorno, S. Di Franco, R. Lo Nigro, F. Giannazzo, P. Prystawko, P. Kruszewski, M. Krysko, E. Grzanka, M. Leszczynski, C. Tudisco, G.G. Condorelli, F. Roccaforte, Phys. Status Solidi A, 212 (2015) 1091-1098
[21] G. Greco, F. Iucolano, C. Bongiorno, F. Giannazzo, M. Krysko, M. Leszczynski, F. Roccaforte, Appl. Surf. Sci. 314 (2014) 546–551. https://doi.org/10.1016/j.apsusc.2014.07.018.
[22] F. Roccaforte, F. Giannazzo, A. Alberti, M. Spera, M. Cannas, I. Cora, B. Pécz, F. Iucolano, G. Greco, Mater. Sci. Semicond. Process. 94 (2019) 164–170. https://doi.org/10.1016/j.mssp.2019.01.036.
[23] R. T. Tung, Mater. Sci. Eng., R. 35,1 (2001).
[24] J. P. Sullivan, R. T. Tung, M. R. Pinto, and W. R. Graham, J. Appl. Phys. 70, 7403 (1991).
[25] R. F. Schmitsdorf, T. U. Kampen, and W. MWnch, J. Vac. Sci. Technol. B 15,1221 (1997)
[26] F. Roccaforte, F. La Via, V. Raineri, R. Pierobon, and E. Zanoni, J. Appl. Phys. 93, 9137 (2003).
[27] F. Roccaforte , G. Greco, P. Fiorenza, S. Di Franco, F. Giannazzo, F. La Via, M. Zielinski, H. Mank, V. Jokubavicius, R. Yakimova, Appl. Surf. Sci. 606 (2022) 154896
[28] G. Greco, P. Fiorenza, M. Spera, F. Giannazzo, F. Roccaforte, J. Appl. Phys. 129 (2021) 234501.
[29] M. Vivona, G. Greco, G. Bellocchi, L. Zumbo, S. Di Franco, M. Saggio, S. Rascuna, F. Roccaforte, J. Phys. D: Appl. Phys. 54 (2021), 055101.
[30] F. A. Padovani and R. Stratton, Solid-State Electron. 9, (1966) 695
[31] H. Kim; J. Electron. Mater. 50, (2021) 6688–6707.